# Particle image velocimetry correlation signal-to-noise ratio metrics and measurement uncertainty quantification

**Zhenyu Xue[1], John J. Charonko[2] and Pavlos P. Vlachos[3]**

[1] Department of Mechanical Engineering, Virginia Tech, Blacksburg, Virginia
conanxzy@vt.edu
[2] Los Alamos National Laboratory, Los Alamos, New Mexico
jcharonk@lanl.gov
[3] School of Mechanical Engineering, Purdue University, West Lafayette, Indiana
pvlachos@purdue.edu

**ABSTRACT**

In particle image velocimetry (PIV) the measurement signal is contained in the recorded intensity of the particle image pattern superimposed on a variety of noise sources. The signal-to-noise-ratio (SNR) strength governs the resulting PIV cross correlation and ultimately the accuracy and uncertainty of the resulting PIV measurement. Hence we posit that correlation SNR metrics calculated from the correlation plane can be used to quantify the quality of the correlation and the resulting uncertainty of an individual measurement. In this paper we present a framework for evaluating the correlation SNR using a set of different metrics, which in turn are used to develop models for uncertainty estimation. The SNR metrics and corresponding models presented herein are expanded to be applicable to both standard and filtered correlations. In addition, the notion of a "valid" measurement is redefined with respect to the correlation peak width in order to be consistent with uncertainty quantification principles and distinct from an "outlier" measurement. Finally the type and significance of the error distribution function is investigated. These advancements lead to robust uncertainty estimation models, which are tested against both synthetic benchmark data as well as actual experimental measurements. For all cases considered here, standard uncertainties are estimated at the 68.5% confidence level while expanded uncertainties are estimated at 95% confidence level. For all cases the resulting calculated coverage factors approximate the expected theoretical values thus demonstrating the applicability of these new models for estimation of uncertainty for individual PIV measurements.

# 1 Introduction

Particle Image Velocimetry (PIV) is a quantitative flow visualization tool developed to measure fluid velocities over a wide range of length and time scales. The technique typically employs flow tracer particles, which are illuminated by a pulsed laser and photographed with a digital camera. Image processing algorithms are then used to estimate the displacement of the particle patterns within an image sequence, and subsequently the velocity field (1). An overview of the development of DPIV over the past twenty years is given by Adrian (2).

PIV was first developed in the 1980s, and the initial work of Meynart (3) was followed by numerous contributions that formally established the foundations of the method (4-7). The introduction of digital image acquisition (1) led to digital PIV (DPIV), which triggered a widespread use and an explosive growth of applications. Over the next twenty years the robustness and accuracy of the technique, including the development of stereoscopic (3-component) planar PIV (8, 9), and iterative, and adaptive methods (10-14) ensued. Currently, the term PIV is used to encompass the extensive family of methods based on evaluating the particle pattern displacement using statistical cross-correlation of consecutive images with high number densities of flow tracers (15).

However, the development of increasingly sophisticated PIV methods far outpaced our ability to quantify the PIV measurement uncertainty. The situation is exacerbated by the fact that PIV measurements involve instrument and algorithm chains with coupled uncertainty sources, rendering quantification of uncertainty more complex than traditional flow measurement techniques. Therefore developing a fundamental methodology for quantifying the uncertainty for PIV is an important and outstanding challenge.

The first attempt to tackle this problem employed an "error-surface" methodology which would be constructed by mapping the effects of selected primary error sources such as shear, displacement, seeding density, and particle diameter to the true error for a given measurement (16). This approach is roughly analogous to a more traditional instrument calibration procedure for standard experimental instruments. The generated error surface provides the means to associate the corresponding distribution of errors to any combination of inputs of the error sources within their parameters space, as quantified directly from the actual experiment. Ultimately in order to comprehensively quantify the uncertainty, all possible combinations of displacements, shears, rotations, particle diameters, and any other parameter used must be tested which can make this method computationally expensive.

Moreover, many of the relevant parameters may not be easily obtained from a real experiment.

Sciacchitano et al. proposed a method to quantify the uncertainty of PIV measurement based on particle image matching (17). The uncertainty of measured displacement is calculated from the ensemble of disparity vectors, which are due to incomplete match between particle pairs, within the interrogation window. This method accounts for random and systematic sources of error; however peak-locking errors and truncation errors cannot be detected. In addition, the disparity can be calculated only for particles that are paired within the interrogation window, this method cannot account for the effects of in plane and out of plane loss of particles and works best for particle image patterns that have been iteratively deformed to converge on each other. Finally, to calculate the instantaneous local uncertainty, researchers need to do particle image pair detection and image matching for every interrogation spot which introduces additional computation cost.

In this work we adopt an alternative approach and seek to quantify PIV measurement uncertainty directly from the information contained within the cross-correlation plane. The cross-correlation plane represents the distribution of probabilities for all possible particle image pattern displacements between consecutive frames, combined with the effect of the number of particles, mean particle diameter and effects that contribute to loss of correlation. In other words, the correlation plane is a surrogate of the combined effects of the various sources of error that govern the estimation of a particle pattern displacement. Hence, in this work we will seek to establish appropriate measures that quantify the cross-correlation quality by means of signal-to-noise ratio (SNR) and establish the relationship between these metrics and the uncertainty of individual measurement.

One measure of the cross-correlation SNR is the primary peak ratio (PPR), namely the ratio between the primary correlation peak to the second tallest peak. In earlier PIV papers, PPR was used as a measure of the detectability of the true displacement (6, 18). A measurement would be considered as valid if PPR were higher than a user defined threshold (often 1.2). Based on this criterion, it was established that the product of $N_I F_I F_O$ value should be approximately 5 (19), where $N_I$ is the particle image density, $F_I$ is the in-plane loss of correlation, $F_O$ is the out of plane loss of correlation. Unfortunately, the effects of in-plane and out-of-plane loss of correlation are difficult to quantify in a real experiment, thus making $N_I F_I F_O$ difficult to estimate in real experiment cases. However this establishes a clear relationship between a measure of the correlation strength (PPR) and number of correlated particle image pairs.

In contrast, the PPR value is easy to compute and provides a practical measure of the quality of a cross-correlation. Hain and Kahler (20) suggest that a threshold PPR value of about 2 can reliably avoid spurious vectors, and based on this they proposed a scheme for the optimal selection of cross-correlations across a range of interframe time delays. Similarly for extending the PIV velocity dynamic range using multiple pulse separation imaging, Persoons and O'Donovan used a weighted peak ratio value as a criterion to calculate the optimum pulse separation (21).

Recently, Charonko and Vlachos proposed an uncertainty quantification method based on PPR (22). The relationship between the distribution of velocity error magnitude and PPR value was studied and a model for calculating the uncertainty based on the PPR value of a given measurement was developed. Using this method, the uncertainty of PIV measurement can be predicted without the a-priori knowledge of image quality and local flow condition. Reliable uncertainty estimation results using a phase-filtered correlation (RPC) (23-25) were shown. However for standard cross-correlation (SCC) techniques, the uncertainty estimation provided by this method was not as reliable. This was attributed to the insufficient treatment of noise effects inherent to the standard cross-correlation. Another problem with this method is that it overestimates the standard uncertainty for both SCC and RPC method. The authors assumed the error distribution of PIV measurement should follow normal distribution, thus the coverage factor should be close to 68.5% for standard uncertainty and 95% for expanded uncertainty. However, this method only succeeded in predicting the expanded uncertainty, the coverage factors were much larger than the theoretical value for standard uncertainty with both phase-filtered correlation and standard cross-correlation.

Beyond the PPR other metrics exist for quantifying the cross-correlation SNR. Kumar and Hassebrook defined several signal to noise ratios of the correlation related to peak detectability including the peak ratio (PPR), namely peak-to-root mean square ratio (PRMSR), and peak-to-correlation energy (PCE) (26). All three of these metrics measure the strength of correlation but the PPR is a mostly heuristic parameter while in contrast the PCE and PRMSR are grounded to signal processing theory. However, within the scope of PIV methods, neither PCE nor PRMSR have previously been considered as measures of correlation quality.

In this work, we extend the original work by Charonko and Vlachos (22) to calculate cross-correlation SNR metrics and develop models for uncertainty estimation. Several contributions are presented. In addition to the PPR, we consider PRMSR, PCE, and cross-correlation entropy (based on information entropy (27)); and we expand the previous work to make these measures applicable to both standard and phase filtered cross-correlation. Moreover, in this process, it was deemed necessary to refine the notion of invalid measurements in order to be consistent with the uncertainty estimation framework. Finally, we investigate the shape of the PIV measurement error distribution since any assumptions inherent therein affect the estimation of the uncertainty confidence intervals. Our method eliminates the assumption that error is normally distributed by directly using the PIV measurement error histogram to build the uncertainty prediction models.

## 2 Background

### 2.1 Correlation plane signal to noise ratio (SNR)

Following the work by Kumar and Hassebrook (26), three measures are introduced to quantify the correlation SNR. These are the primary peak ratio (PPR), the peak to root mean square ratio (PRMSR) and the peak to correlation energy (PCE).

The primary peak ratio (PPR) is defined as the ratio between the height of the primary peak and the height of the second tallest peak, it is calculated as:

$$PPR = \frac{C_{max}}{C_2} \qquad (1)$$

The signal part is $C_{max}$, the primary peak height, and the noise part is $C_2$, the height of the secondary peak.

The peak to root mean square ratio (PRMSR) is defined as the ratio between the magnitude of cross correlation plane and square of the correlation plane root mean square value. Expression for PRMSR calculation is:

$$PRMSR = \frac{|C_{max}|^2}{C_{rms}^2} \qquad (2)$$

Signal part is the magnitude of the cross correlation plane $|C_{max}|^2$, which is calculated as the square of the primary peak height; the noise part $C_{rms}^2$, and is calculated as:

$$C_{rms}^2 = \left[\sqrt{\frac{1}{N_\Omega} \sum_{i \in \Omega} |C(i)|^2}\right]^2 \qquad (3)$$

where $\Omega$ indicates the group of points on the correlation plane where the correlation value of those points are lower than half of primary peak height.

The peak to correlation energy (PCE) is defined as the ratio between the magnitude of cross correlation plane and the correlation energy, which can be calculated as:

$$PCE = \frac{|C_{max}|^2}{E_c} \quad (4)$$

The magnitude of the correlation plane $|C_{max}|^2$ is the signal part; the noise part, correlation energy is defined as:

$$E_c = \int_{-\infty}^{\infty} |C(x)|^2 dx \quad (5)$$

However, in practice, the correlation plane has finite size. So we calculate the correlation energy as:

$$E_c = \frac{1}{W}\left(\sum_W |C(x)|^2\right) \quad (6)$$

where W is the size of the correlation plane.

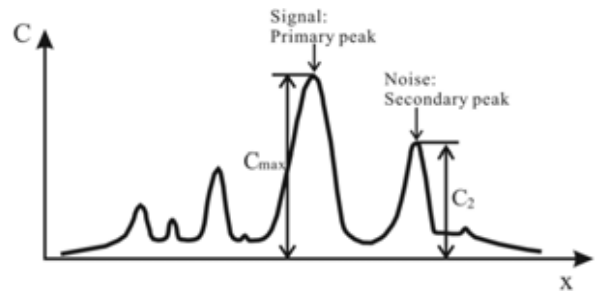

(a)

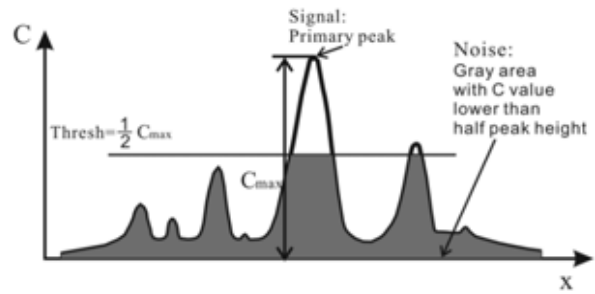

(b)

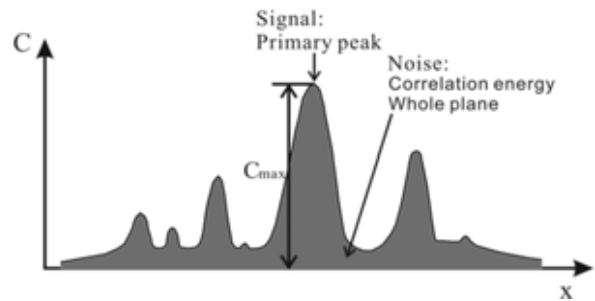

(c)

**Figure 1** 1D Graphical representation of correlation SNR: (a) PPR; (b) PRMSR; (C) PCE.

Figure 1 shows one-dimensional graphical representation of all three SNR metrics. Effectively these metrics measure the detectability of the primary peak with respect to alternative correlations. However, in contrast to the PPR which is an *ad-hoc* metric, the PRMSR and PCE are amenable to analytical derivation if the signal statistical properties are known (26), hence they offer the potential for developing a corresponding theoretical foundation for the uncertainty estimation. This aspect however will not be pursued during this work.

Another signal to noise ratio measure considered herein is the cross-correlation entropy or information entropy (27). This is based on the notion that if perfect matching between two image patterns exists in the absence of any noise, the correlation will yield a single sharp peak and the correlation entropy value will be minimum. As more random correlations exist the entropy would increase. To calculate the entropy of the cross correlation plane, we first construct the histogram of the correlation plane based on the correlation value of every point on the plane. In our work, we use 30 bins to build the histogram. After the histogram is generated, the probability of finding one point within a certain bin is calculated as:

$$p_i = \frac{\text{\# of points @ bin } i}{\text{Total \# points of whole plane}}$$

Then the entropy of the cross correlation plane was calculated as:

$$Entropy = \sum_{i=1}^{30} p_i \log \frac{1}{p_i} = -\sum_{i=1}^{30} p_i \log p_i \qquad (7)$$

## 2.2 Role of image background noise on correlation SNR

The information about the true displacement in the correlation plane is contained in the correlation of the fluctuating intensities. If the correlation is written as:

$$R(s,t) = R_C(s,t) + R_D(s,t) + R_F(s,t) \qquad (8)$$

where the overall correlation plane is decomposed into $R_C$, $R_F$, and $R_D$, which are respectively the correlation of the mean background intensity over the interrogation windows, the correlation of the background noise in one window with the fluctuating intensity in the other window, and the cross correlation of the fluctuating image intensities. It is common practice to subtract the image mean intensity before performing a cross-correlation, which would effectively remove all contributions from the background and only provide $R_D$. However in practice this does not always hold true due to various illumination artifacts and imaging distortions. Although for the estimation of the true displacement such residuals would have negligible effect, in contrast for the calculation of the correlation SNR they can profoundly affect the metrics. The reason for this can be simply illustrated by the fact that if any two arbitrary measures are increased by the same constant offset, although their difference remains the same, their ratio is decreasing. By definition the SNR is a ratio of the measure of the signal to the noise levels hence would be subject to the same effect. As a result, in their work Charonko and Vlachos (22) showed that the standard correlation suffered from this effect and performed inferiorly to the phase filter correlation which in turn is largely immune to such effects. Hence in order to address this limitation and provide more robust estimation of the different correlation metrics we propose to subtract the minimum value of the correlation plane.

Figure 2 demonstrates this using an example of a particle image with and without background noise. The corresponding cross correlation planes of these two image sets are shown in Figure 3 a and b. The minimum correlation value of the cross correlation is on the order of $10^6$. After we subtract the correlation plane of Figure 3b ($R_D$) from Figure 3a ($R$), the left plane Figure 3c can be considered as the correlation related to background image noise, the $R_C$ and $R_F$ term together. The mean value of this plane is also close to $10^6$. Subtraction of the minimum correlation value from the correlation plane effectively eliminated the effect of background image noise on the cross correlation plane without having to explicitly filter the image regions.

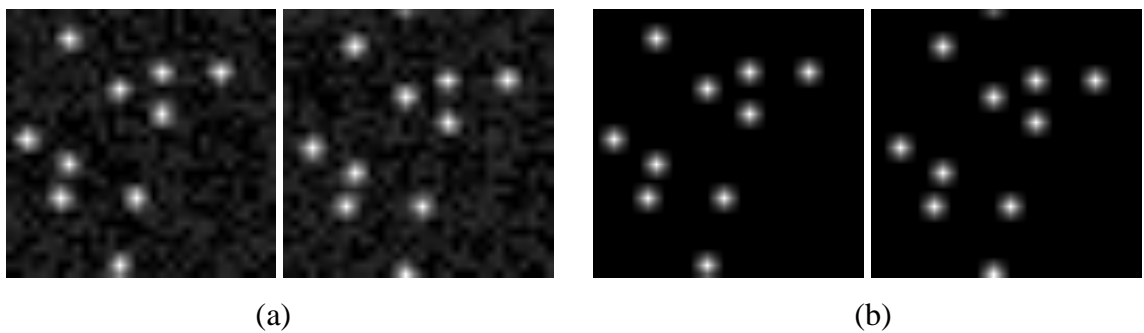

(a)                                                              (b)

**Figure 2** Particle image sets examples (a) with background noise; (b) same particle images without background noise.

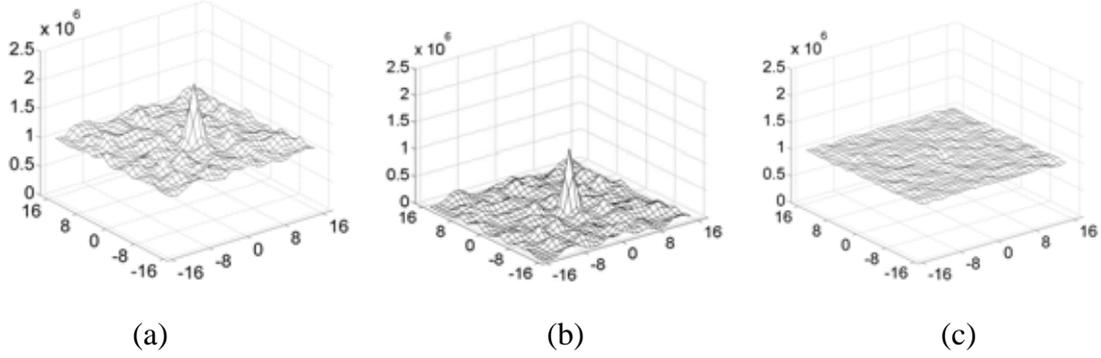

(a)                 (b)                 (c)

**Figure 3** (a) cross correlation plane of particle images with background noise (b) cross correlation plane of particle images without background noise (c) the correlation plane related to background noise.

## 2.3  Correlation width and definition of a valid measurement

The primary peak diameter can be calculated by:

$$d_D = \sqrt{2d_\tau^2 + \frac{4}{3}a^2} \qquad (9)$$

where $d_\tau$ is the particle image diameter and a is a velocity gradient parameter[15]. However, for a given correlation plane, the correlation peak width is usually calculated by performing a three-point Gaussian fit and then computing the diameter as 4 times the standard deviation for the Gaussian distribution (28). Using an elliptical Gaussian fit would provide a better estimation of the size for the cases when the correlation does not yield a symmetric Gaussian distribution. The location of the maximum value of that Gaussian function provides the sub-pixel displacement estimation for the PIV measurement. This is subject to the assumption that the true displacement is within the primary peak region. Thus, if the difference between the measured displacement and true displacement (i.e. the error) is less than half of the peak diameter, the measurement should be considered as valid because the peak corresponds to the true displacement. However, previous works often use a fixed threshold value for detecting failed measurements or outliers. Outliers were identified either using a local neighbourhood statistical criterion(29) or if the difference between the measurement and the true value is larger than a pre-determined threshold, for example 0.5 or 1 pixel, regardless if the correlation peak contains the true displacement information or not. By using either approach, the conventional definition of outliers is inconsistent with the notion of error and uncertainty. For example, a wide peak at a location corresponding to the true displacement, although it could yield errors in excess of 1 pixel, would still be accurate but it would not be precise.

This is because the identified displacement peak successfully matched the true measurement, but was too wide to yield a precise evaluation of the displacement. Instead, here we suggest that the criterion for a valid measurement should be based on the diameter of the correlation peak. If the error is less than half of the peak diameter, we conclude that the measurement successfully found the correct peak and it is indeed a valid measurement. Only those measurements providing the wrong primary peak are considered as invalid. An example of this "half peak diameter" criterion is shown in Figure 4.

Note that the concepts of valid measurements versus outliers are different and distinct. An outlier is determined by statistical comparison with its neighbourhood while a valid or invalid measurement should be based on an independent assessment of the measurement's success or failure, regardless of the statistical properties of the neighbourhood in which it is located. Using this model, a peak with diameter larger than a single pixel, due to particle size and large shear gradient, may be identifying the velocity distribution within the interrogation region correctly; this holds even if the highest point within the peak is located more than one pixel away from the velocity value at a location in the center of the correlation region. Thus, it should not be counted as a failure, but should instead be included as a valid measurement but with a larger uncertainty.

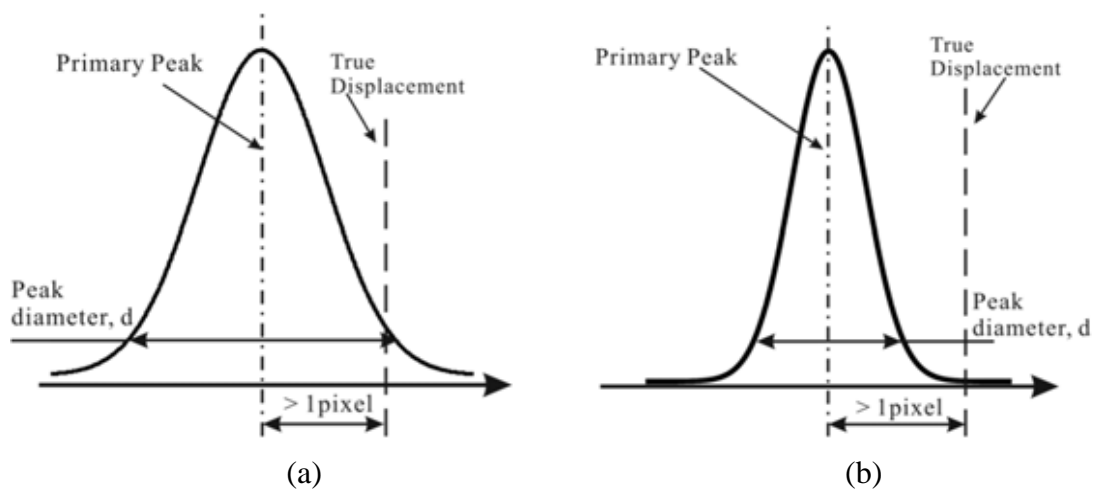

**Figure 4** 1-D example of half peak diameter criterion (a) good measurement; (b) outlier.

## 2.4 Synthetic image sets

Synthetic image sets with known displacements information were used to develop the relations between the uncertainty or error distribution and the measured metrics' value.

### 2.4.1 Turbulent boundary layer

The first data set is 100 image pairs of turbulent boundary layer flow field (Case B of the Second International PIV challenge in 2003 (30)). The images contain 70 particle pairs per 32X32 region with 2.6 pixel average particle diameter at 4 standard deviations.

### 2.4.2 Laminar separation bubble

The second data set is 18 image pairs of laminar separation bubble flow field (Case B of the Third International PIV challenge in 2005 (31)). 25 particles per 32X32 window is the average seeding density of this data set. The average particle diameter is approximately 2.0 pixels. Only six timesteps (frames 10, 30, 50, 70, 90 and 110) of exact solution of flow field were provided. Particle images of symmetrically distributed frames about the desired timestep were correlated, for example, the flow field calculated by correlating frame 9 and 11 is compared with the provided exact solution of frame 10. The maximum correlation step in our test is 6 frames (for example, frame 7 correlated with frame 13).

### 2.5 Statistical analysis, error distribution and uncertainty estimation

We follow the same steps to calculate the error of each measured vector described by Charonko and Vlachos [22]. After we calculate the values of the metrics mentioned before and the error of all the vectors in the two synthetic image data sets, we divided all the data points into 40 bins based on the value of the calculated metrics. Charonko and Vlachos have shown that the difference between the absolute magnitudes of mean velocity error and absolute mean error plus the standard deviation was very small (22), so therefore they used $rms\delta_v$ for the error distribution, calculated as:

$$rms\delta_{v,i} = \sqrt{\left(mean\left(\delta_{v,i}^2\right)\right)} = \sqrt{\frac{1}{N}\sum_{i=1}^{N}\delta_{v,i}^2} \qquad (10)$$

They also assumed that the error distribution of PIV measurement follows normal (Gaussian) distribution. Thus the coverage factor of standard uncertainty should reflect the probability that the true error stays within one standard deviation range. For normal distribution, the theoretical value for one standard deviation is 68.5%. However, this assumption was not tested at the time. Moreover in their analysis the standard uncertainty range was consistently over predicted with values of 81.4% for SCC and 76.1% for RPC, suggesting that the error distribution is likely not Gaussian.

In this work, we now test the validity of this hypothesis carefully analysing the error distribution type of PIV measurements. For example, Figure 5a shows the histogram of SCC error magnitude in the bin where PPR value is 1.5; 26727 data points were included in this bin. The solid line shows a Gaussian function based on the model Charonko and Vlachos used [22], where the mean value is fixed at 0 and the standard deviation value is the root-mean-square value of the error magnitude in of this bin. The dash line is the Gaussian function curve fitting based on the histogram and the dash-dot line represents the lognormal function curve fitting based on the histogram. More details can be seen when zoomed in, as shown in Figure 5b. It is obvious that the data do not follow a Gaussian (normal) distribution and a lognormal distribution might provide a better approximation.

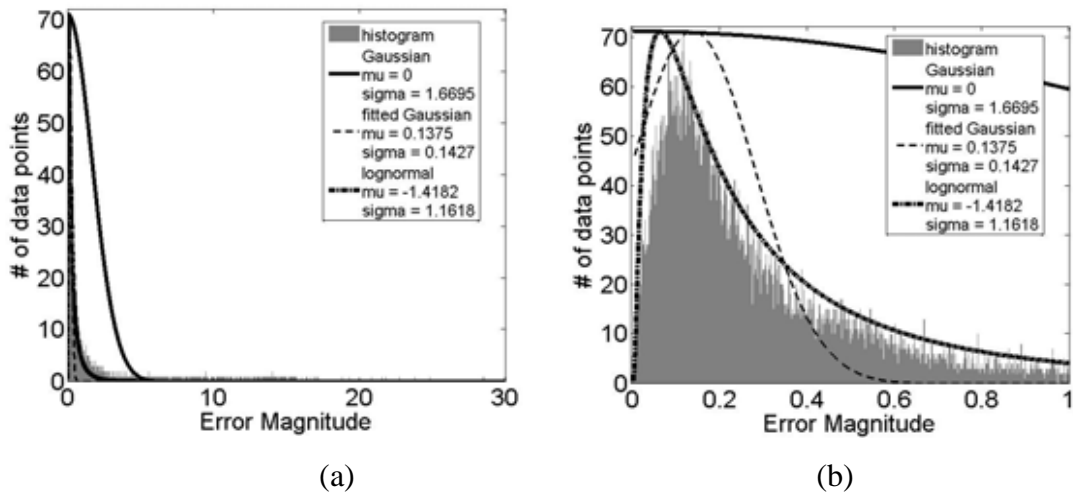

**Figure 5** Histogram and different distribution fitting of error magnitude at PPR = 1.5 (a) whole range of error; (b) zoomed in error < 1 region.

We did the same test as shown above across all the range of available data. A few examples of histogram and the fitted curves error distribution function are shown in Figure 6. It is clear that across the whole range of PPR value, the lognormal fitting provides a better performance than Gaussian fitting especially at the tail regions. However, the lognormal fitting is still not a good match to the histogram at the tail region. As shown in all figures in Figure 6, the lognormal fitting curve is still lower than the outline of the histogram in the tail region. Other types of distributions were also checked and across all the range of available data, and lognormal fitting was the best among all the tested error distribution functions. By checking the histogram of PIV error magnitude, two important facts were discovered: First, the error distribution is not symmetric, suggesting that we should include two different parameters for upper and lower limit of the uncertainty instead of using only one parameter as in previous work (22); Second, it is actually improper to assume that the distribution of PIV error magnitude must follow a certain predefined function.

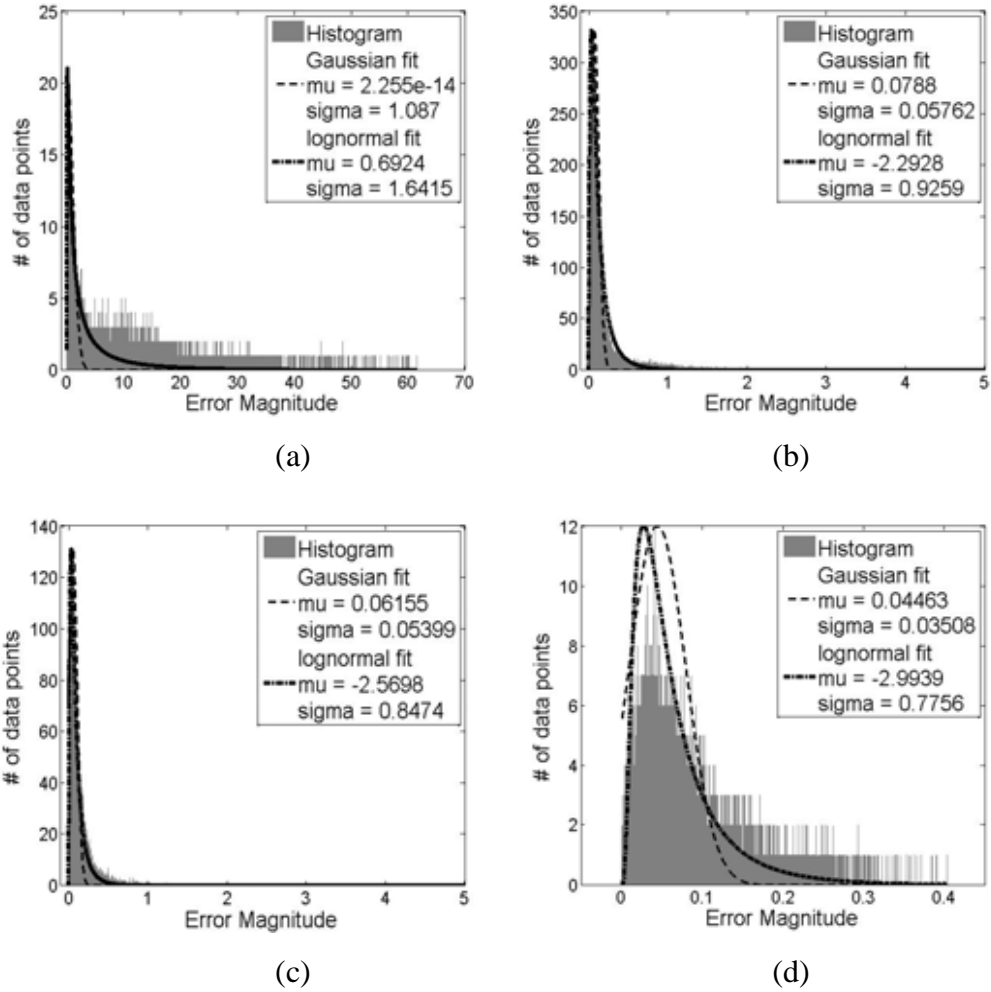

**Figure 6** Histogram and different distribution fitting of error magnitude at (a) PPR = 1.0; (b) PPR = 3.0; (c) PPR = 5.9; (d) PPR = 9.9.

Therefore, for developing models for PIV uncertainty estimation, instead of using the value of error magnitude, we calculate the upper and lower limit of the uncertainty based on the histogram of the binning as described in the previous paragraph. We keep using 68.5% and 95% confidence level for standard and expanded uncertainty, which means 68.5% and 95% of all the data have the error magnitude smaller than the predicted uncertainty calculated by our model. Figure 7 is an example of calculating the upper and lower limit for expanded uncertainty of the bin with PPR value of 9.9. As shown in the figure, $e_{low}$ is the lower limit indicating the point where the area under the curve is 2.5% of total area under the error magnitude histogram; $e_{high}$ is the upper limit and 97.5% of the total area is under the curve on the error magnitude histogram. For standard uncertainty, the ratio is changed to 15.75% and 84.25% for lower and upper limit respectively. This approach allows robust estimation of the uncertainty coverage without any a-priori assumption of an error distribution function.

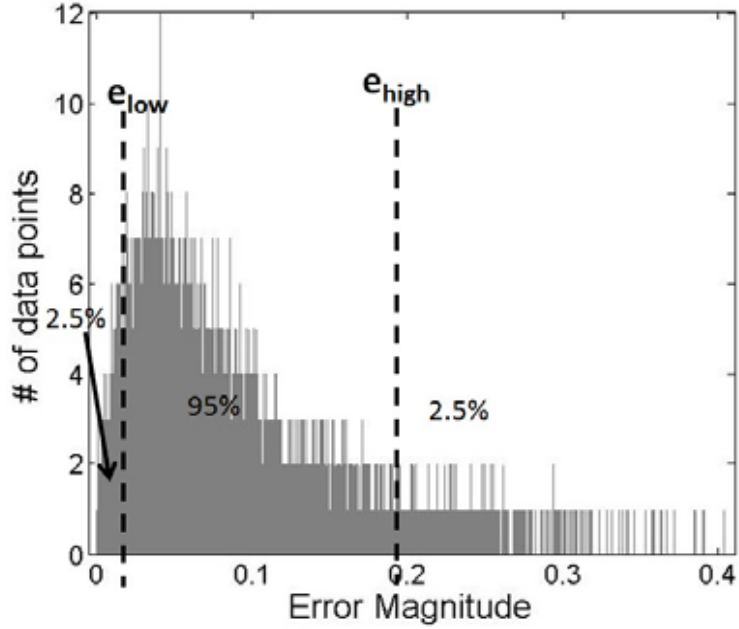

**Figure 7** Example of calculating upper and lower limit for expanded uncertainty.

### 3 Result and discussion

#### 3.1 Relationship of uncertainty versus cross-correlation SNR metrics

The uncertainty model which provides a relationship between any of the SNR metrics to the standard and expanded uncertainty limits are based on a fitting function proposed by Charonko and Vlachos (22). Here, the estimated both standard uncertainty (u) and expanded uncertainty (U) are calculated by determining the fitting parameters of the following equation:

$$u^2 = \left( M \exp\left( -\frac{1}{2}\left( \frac{\phi - N}{s} \right)^2 \right) \right)^2 + \left( A\phi^{-B} \right)^2 + C^2 \tag{11}$$

$$U^2 = \left( M \exp\left( -\frac{1}{2}\left( \frac{\phi - N}{s} \right)^2 \right) \right)^2 + \left( A\phi^{-B} \right)^2 + C^2 \tag{12}$$

The first term is a Gaussian function used to account for the uncertainty due to invalid measurements which contribute uncertainty M, where the exact value of M is related to the range of possible velocity measurements and the distribution of the true velocity within the sampled flow field (22). The $(\phi - N)$ term allows the error to climb rapidly as the metric's value approaches some small number, and N is the theoretical minimum value of the calculated metric.

Based on the definition of each quantity, we can determine analytically what value of N we should use for each metric. For PPR, the minimum value is Min(PPR) = 1 when we have a primary peak and secondary peak with the same height. Based on the definition of PRMSR, when all points in $C_{rms}$ have a value of half the main peak height, the theoretical minimum value for PRMSR is Min(PRMSR) = 4. The extreme case for PCE occurs when the peak is only slightly higher than the rest of the correlation plane, and the rest of the plane shared the same correlation value; in this case the minimum PCE value Min(PCE) approaches 1. Because entropy behaves the opposite way as the other basic SNR metrics, we take the inverse of entropy, i.e. $\phi = 1/entropy$ to keep the fitting function type consistent among all metrics. Therefore, the theoretical minimum value for inverse entropy should be 0 as entropy approaches infinity.

The second power-law term in equations 11 and 12 is the contribution to the uncertainty by the valid vectors, which means the largest uncertainty that could be expected would be governed primarily by A if it can be assured that the given measurement is valid. The last term C is a constant, which corresponds to the lowest uncertainty we can achieve. The estimated uncertainty for a measurement with a given calculated metric value is governed by the combination of the above three terms.

Although invalid vectors are detected by using the new half peak diameter rule described earlier, it is not appropriate to develop a model for uncertainty estimation using only the valid measurements. Unlike synthetic image sets we used to build the model, in real experiments the true velocity field is unknown and it is inevitable that velocity fields would be contaminated by invalid measurements. Therefore both invalid and valid vectors are included in developing the uncertainty model estimation. All synthetic cases with 3 different window size (or effective window size for RPC method), 16x16, 32x32, and 64x64 were included in the test providing a sample size containing about 3 million data points in total.

### 3.2 Result of uncertainty estimation

In order to keep the calculation process consistent, we applied the minimum correlation subtraction method as described earlier to both SCC and RPC. However, this method has a minimal effect on the RPC models. Figure 8 shows the curve fitting result for estimation the uncertainty using peak ratio with both RPC and SCC methods after the minimum subtraction.

The resulting curve fitting parameters for SCC and RPC are shown in **Table 1**. Where UL stands for upper limit and LL stands for lower limit.

**Table 1** Fitting parameters of PPR.

|  | PPR | | | | | | | |
|---|---|---|---|---|---|---|---|---|
|  | SCC | | | | RPC | | | |
|  | $u_{UL}$ | $u_{LL}$ | $U_{UL}$ | $U_{LL}$ | $u_{UL}$ | $u_{LL}$ | $U_{UL}$ | $U_{LL}$ |
| M | 10.59 | 0.278 | 23.69 | 0.06816 | 25.11 | 0.09828 | 28.33 | 0.03258 |
| N | 1 | 1 | 1 | 1 | 1 | 1 | 1 | 1 |
| S | 0.1925 | 0.1927 | 0.2753 | 0.2446 | 0.2874 | 0.2258 | 0.7188 | 0.2406 |
| A | 0.6888 | 0.1043 | 4.112 | 0.03619 | 0.4583 | 0.09359 | 2.32 | 0.0359 |
| B | 0.846 | 0.6786 | 1.357 | 0.6342 | 0.5696 | 0.5597 | 0.8399 | 0.5681 |
| C | 0 | 0 | 0 | 0 | 0 | 0 | 0 | 0 |
| $R^2$ | 0.98 | 0.97 | 0.98 | 0.97 | 0.97 | 0.95 | 0.98 | 0.94 |

In previously reported result for SCC processing, the fitting curve only partially agreed with the original data (22). The current results shown in Figure 8a, show that the fit model provides agreement with the original data almost across the whole range, with R square value of 0.98 for the upper limit of standard uncertainty, 0.97 for the lower limit of standard uncertainty, 0.98 for the upper limit of expanded uncertainty and 0.97 for the lower limit of expanded uncertainty. The model for estimating uncertainty for SCC processing provides larger values by comparison to the RPC processing, which is anticipated since RPC has been shown to yield measurements with lower error. In the end, the current relationships yield an improved fit compared to those previously reported which did not use minimum subtraction of the correlation plane.

The resulting fitting parameters of other metrics are listed in Table 2. Figure 9 shows the curve fitting results for estimating the uncertainty using PRMSR and Figure 10 shows the results using entropy for both SCC and RPC methods. The fitting figures of PCE are similar to PRMSR, so they are not shown in this paper.

All the fit functions agree that the uncertainty would be larger for the SCC than the RPC for the same value of each metric. All these functions showed good agreement with the raw data, and the corresponding R square values are larger than 0.9.

Finally, the percent coverage of the standard and expanded uncertainty was calculated in comparison to the exact true error for each velocity vector measurement according to the following formula:

$$\text{coverage} = \frac{u_{LL}(\text{or} U_{LL}) \leq \text{\# of estimateds for which } \delta_V \leq u_{UL}(\text{or} U_{UL})}{\text{Total \# of velocity estimates}} \times 100\% \tag{13}$$

The coverage should be close to 68.5% for the standard uncertainty and 95% for the expanded uncertainty if the uncertainty estimation was correct on average. The exact value of coverage factor of all functions using all synthetic data sets with 16x16, 32x32 and 64x64 window sizes are shown in Figure 11. In the work by Charonko and Vlachos (22), the authors showed that their method could only match the expanded uncertainty, it overestimated the standard uncertainty and the coverage factors were often greater than the theoretically expected value. Their results also showed that the performance of SCC was inferior than that of RPC. For the present results, the coverage factors are much closer to the theoretical value for both standard uncertainty and expanded uncertainty with both SCC and RPC method. At standard uncertainty level, the SCC peak ratio (PPR) model provides the best match with a resulting coverage value of 68.27%; the worst cases is the SCC entropy model which has a coverage value of 66.58%. At expanded uncertainty level, the best match is the SCC peak ratio (PPR) model with resulting coverage factor of 94.96%. Even for the worst case, the RPC entropy model, the difference between the result and the theoretical value is smaller than 1%.

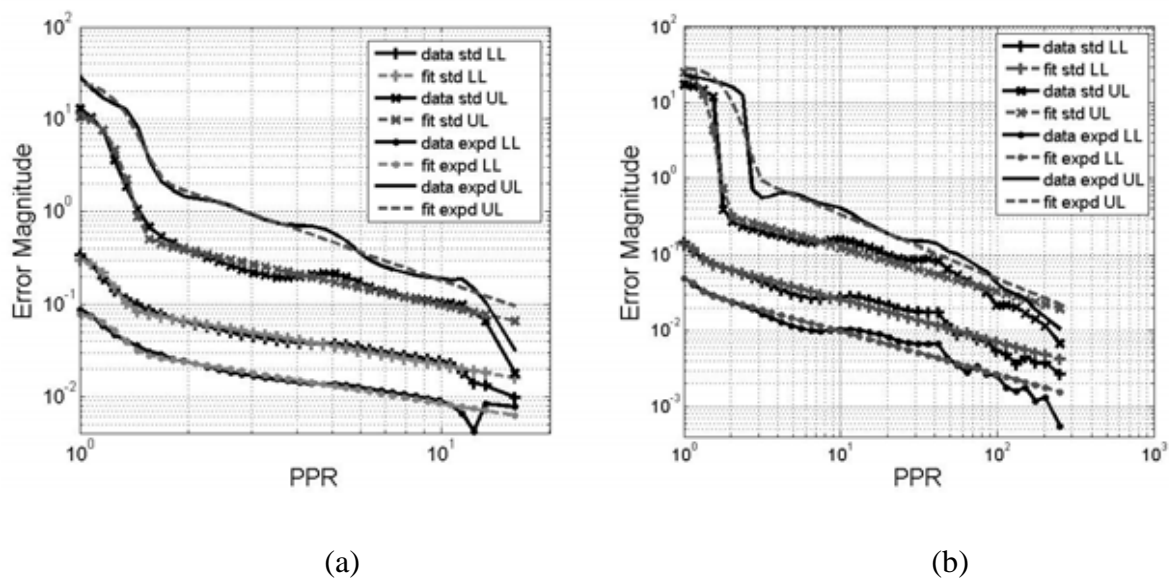

(a)          (b)

**Figure 8** Plots of the relationship of the calculated uncertainty on velocity versus peak ratio for both (a) SCC, and (b) RPC, for all synthetic image sets. (solid lines) original curve of uncertainty bounds versus peak ratio; (dash lines) three term function fitted curve.

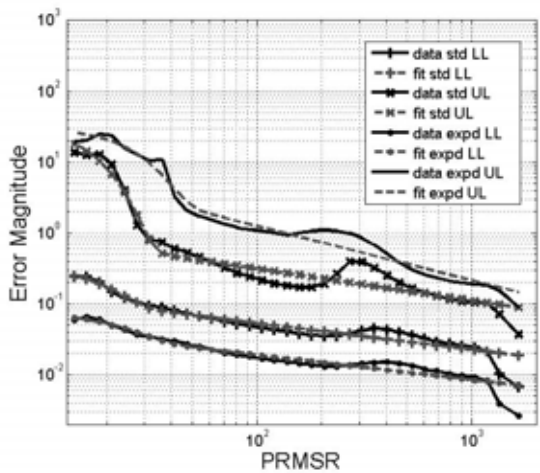 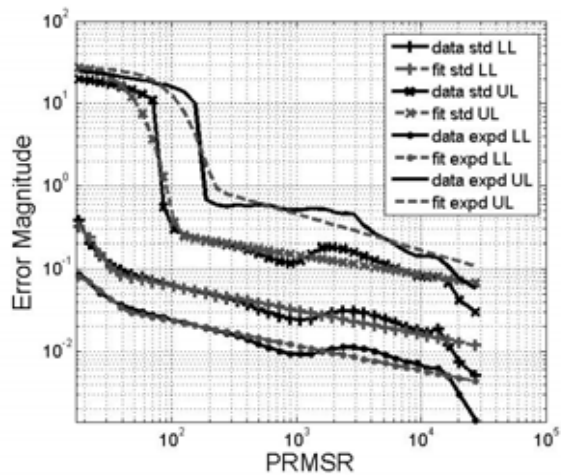

(a)           (b)

**Figure 9** Plots of the relationship of the calculated uncertainty on velocity versus PRMSR for both (a) SCC, and (b) RPC, for all synthetic image sets. (solid lines) original curve of uncertainty bounds versus peak ratio; (dash lines) three term function fitted curve.

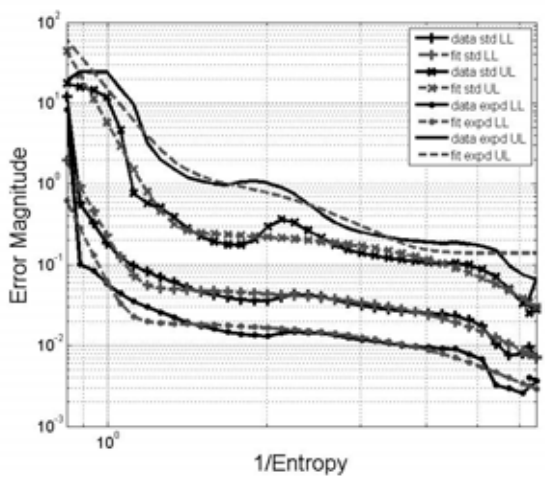 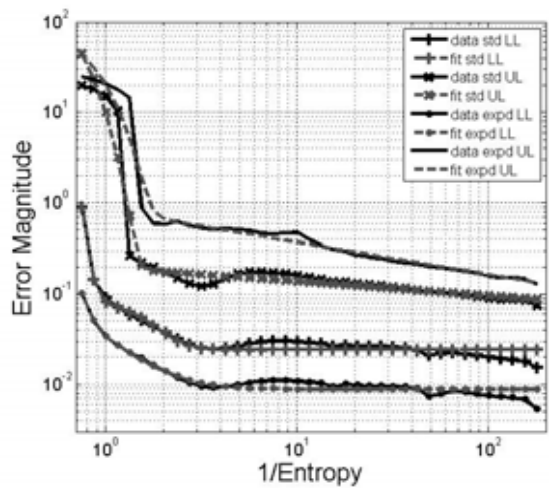

(a)           (b)

**Figure 10** Plots of the relationship of the calculated uncertainty on velocity versus 1/Entropy for both (a) SCC, and (b) RPC, for all synthetic image sets. (solid lines) original curve of uncertainty bounds versus peak ratio; (dash lines) three term function fitted curve.

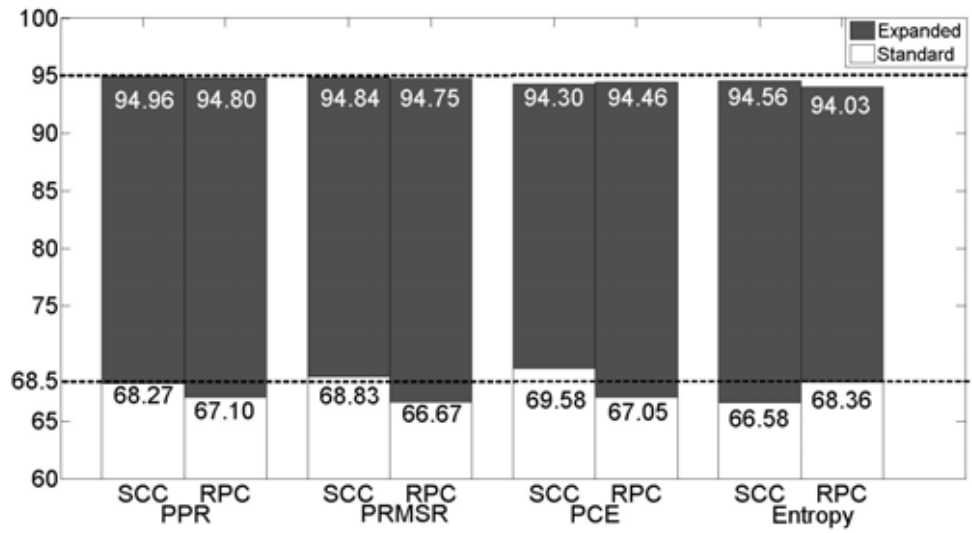

**Figure 11** Coverage factors with synthetic image sets.

**Table 2** Fitting parameters of PRMSR, PCE and Entropy.

|   | PRMSR | | | | | | | |
|---|---|---|---|---|---|---|---|---|
|   | SCC | | | | RPC | | | |
|   | $u_{UL}$ | $u_{LL}$ | $U_{UL}$ | $U_{LL}$ | $u_{UL}$ | $u_{LL}$ | $U_{UL}$ | $U_{LL}$ |
| M | 30.81 | 0.304 | 30.5 | 0.05974 | 29.93 | 0.5268 | 27.45 | 0.09569 |
| N | 4 | 4 | 4 | 4 | 4 | 4 | 4 | 4 |
| S | 9.806 | 13.04 | 17.8 | 18.45 | 32.49 | 13.23 | 76.7 | 17.92 |
| A | 2.494 | 0.2928 | 41.98 | 0.09639 | 0.7614 | 0.2461 | 8.852 | 0.09377 |
| B | 0.4498 | 0.3712 | 0.7634 | 0.3516 | 0.2369 | 0.297 | 0.4312 | 0.3006 |
| C | 0 | 0 | 0 | 0 | 0 | 0 | 0 | 0 |
| $R^2$ | 0.95 | 0.91 | 0.97 | 0.91 | 0.97 | 0.93 | 0.97 | 0.92 |
|   | PCE | | | | | | | |
|   | SCC | | | | RPC | | | |
|   | $u_{UL}$ | $u_{LL}$ | $U_{UL}$ | $U_{LL}$ | $u_{UL}$ | $u_{LL}$ | $U_{UL}$ | $U_{LL}$ |
| M | 23.04 | 6.027 | 30.74 | 0.2045 | 30.28 | 37.77 | 29.38 | 0.1044 |
| N | 1 | 1 | 1 | 1 | 1 | 1 | 1 | 1 |
| S | 7.565 | 2.965 | 11.42 | 0.5224 | 23.79 | 3.837 | 47.76 | 14.29 |
| A | 1.639 | 1.108 | 29.08 | 0.5224 | 0.6188 | 0.7006 | 4.204 | 0.09491 |
| B | 0.3989 | 0.7855 | 0.7533 | 0.8713 | 0.2203 | 0.5618 | 0.3569 | 0.3268 |
| C | 0 | 0.0279 | 0 | 0 | 0 | 0.02047 | 0 | 0 |
| $R^2$ | 0.95 | 0.94 | 0.97 | 0.92 | 0.76 | 0.92 | 0.94 | 0.90 |
|   | Entropy | | | | | | | |
|   | SCC | | | | RPC | | | |
|   | $u_{UL}$ | $u_{LL}$ | $U_{UL}$ | $U_{LL}$ | $u_{UL}$ | $u_{LL}$ | $U_{UL}$ | $U_{LL}$ |
| M | 0.2739 | 0.05352 | 1.757 | 0.02047 | 307 | 0.1039 | 118.3 | 2.181 |
| N | 0 | 0 | 0 | 0 | 0 | 0 | 0 | 0 |
| S | 3.076 | 3.188 | 3.056 | 3.063 | 0.3834 | 1.258 | 0.5364 | 0.2975 |
| A | 5.527 | 0.02148 | 14.7 | 0.05605 | 0.1982 | 0.01667 | 0.8363 | 0.03248 |
| B | -11.59 | -12.25 | -7.985 | -13.41 | -0.1566 | -13.92 | -0.3523 | -1.528 |
| C | 0 | 0 | 0.1407 | 0.001905 | 0 | 0.02452 | 0 | 0.008889 |
| $R^2$ | 0.96 | 0.93 | 0.86 | 0.96 | 0.95 | 0.94 | 0.97 | 0.93 |

## 3.3 Application to experimentally measured flow fields

Our uncertainty models were further tested with real experiment data. In this work, we are using the same data set of stagnation plate flow used by Charonko and Vlachos (22). The experiment details can be found therein. The details of calculating the time average field and the uncertainty introduced by the fitting process were also described in (22).

The exact values of coverage factor of each metric using the real experimental dataset with 32x32 window sizes are shown in Figure 12. The SCC peak to root-mean-square ratio (PRMSR) model performs the best for both standard uncertainty and expanded uncertainty; the resulting coverage values are almost identical to the theoretical values, 68.51% for standard uncertainty and 95.03% for expanded uncertainty. At standard uncertainty level, the worst case is RPC peak ratio (PPR) model, which has the coverage factor of 63.26%. At expanded uncertainty level, the RPC peak to correlation energy (PCE) model performs the worst, it overestimated by 2.8%. The differences between the calculated values and theoretical coverages stay within less than 5% for all cases.

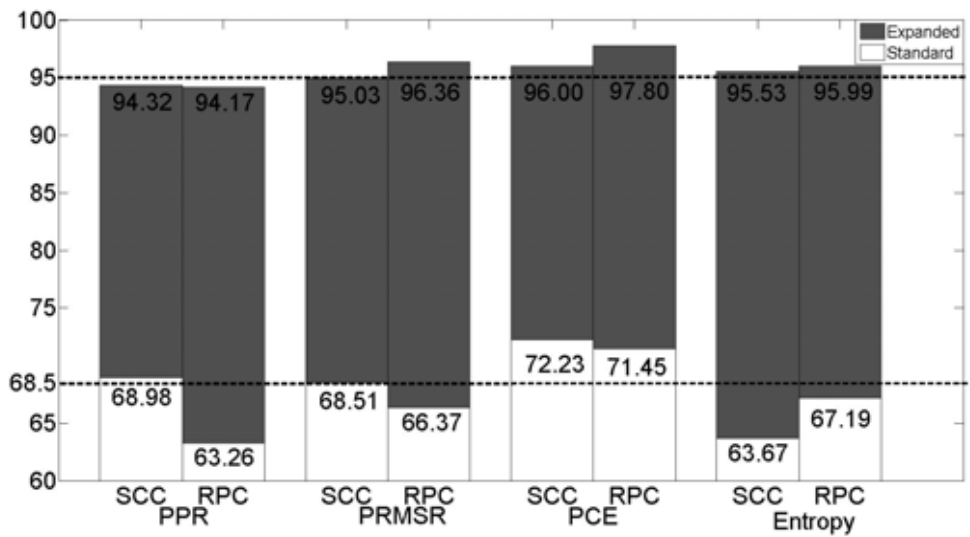

**Figure 12** Coverage factors from the experimental data set.

Figure 13(a) shows the magnitude of the error on a representative instantaneous flow field, processed using SCC method. The peak ratio for each vector in this frame is shown in Figure 13(b), and the reverse pattern is seen. The upper limit of the standard uncertainty calculated from our SCC PPR model of each measured vector is plotted in Figure 13(c), the pattern of the instantaneous uncertainty is quite similar to the error magnitude (Figure 13(a)). However, the magnitude of predicted uncertainty is larger than the error magnitude. This is because in Figure 13(c) what we plot is the upper limit of standard uncertainty, which means statistically about 84% of the error will be less than the value we predict. In fact, about 90% of error estimates are less than the uncertainty value we calculated from our model. The resulting coverage is shown in Figure 13(d), the pixel will be set in color gray if the error magnitude of the measured vector at that location stays within the predicted upper and lower limit range; if the error magnitude is out of the predicted range, it will be colored in black. The total coverage factor for this frame is 62.46%. However, although this estimate only includes 64x80 (5120) data points, the difference between the calculated coverage and the expectation stays within 10%, and when the analysis is expanded to the full data set the agreement is improved, as previously shown in Figure 12.

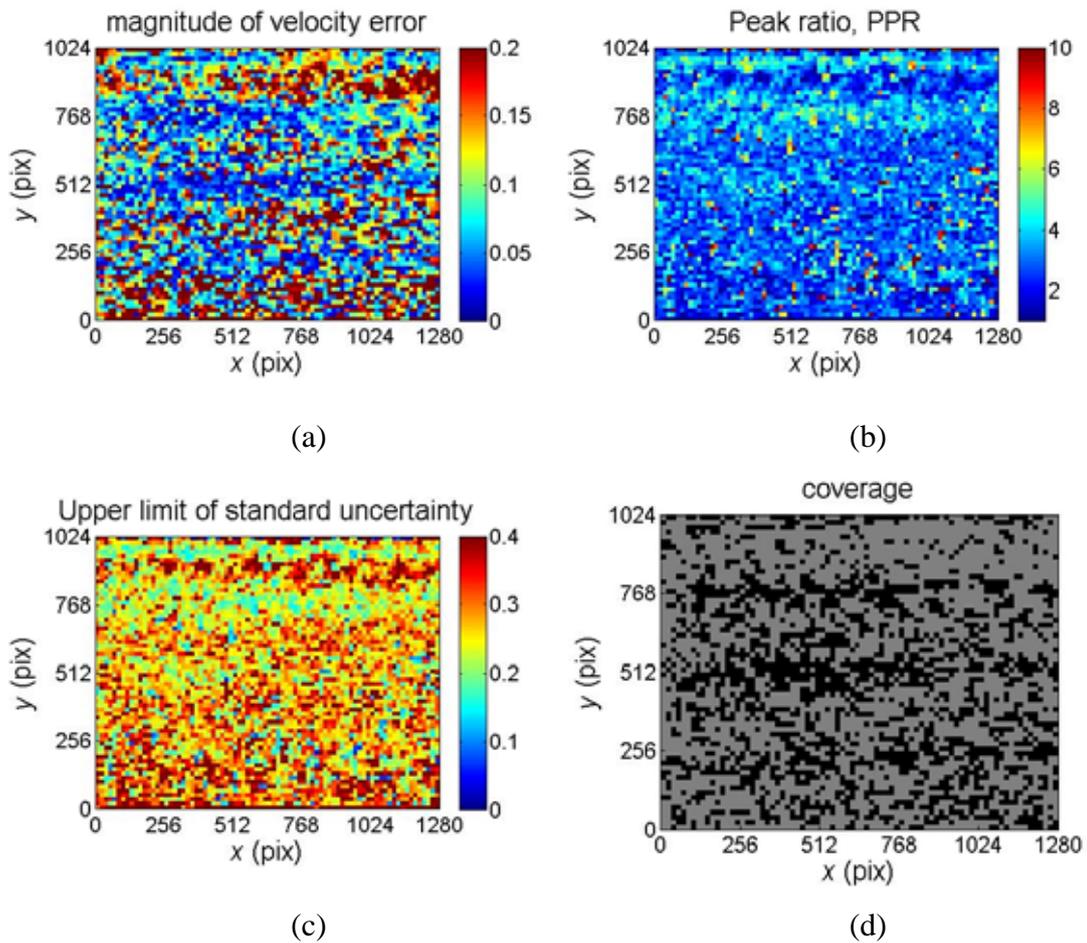

(a)            (b)

(c)            (d)

**Figure 13** (a) Magnitude of estimated displacement error for a representative instantaneous flow field. (b) Peak rations, PPR for the same frame. (c) Upper limit of standard uncertainty for the same frame calculated by SCC PPR model. (d) Coverage ($u_{LL}$ < Magnitude of error < $u_{UL}$), gray indicates the error of the measured vector stays within the estimated uncertainty range while black indicated a failure in estimate the uncertainty bounds.

## 4  Conclusions

In this paper, we show that general relationships exist between cross-correlation SNR metrics calculated exclusively from the correlation plane and that these can be used to develop predictive models for estimating the uncertainty of the PIV measurements. In the first part of our work, metrics of basic correlation SNR related to the peak detectability are presented. A simple but consequential correction on the correlation plane is introduced using a subtraction of the minimum correlation value to remove the effect of the background image noise and thus improve the model's performance for uncertainty estimation. Moreover we redefine the concept of invalid measurement using the correlation peak diameter in order to formulate a definition that is consistent with the uncertainty estimation framework. Finally, in the process of developing uncertainty prediction models we also show that the PIV measurement error distribution cannot be simply assumed to follow a certain known distribution function, and new models with uncertainty upper and lower limit calculated from the error histogram are established, thus first correcting for overestimation errors introduced by a Gaussian distribution assumption but also enabling the development of explicit expressions for the upper and lower limit of standard and expanded uncertainty in our models.

The relationship between the uncertainty and the metrics of correlation SNR of individual velocity measurements were explored using both robust phase correlation (RPC) and standard cross correlation (SCC) processing. The uncertainty is governed by a well-defined relationship between the correlation SNR using a three-term formulation for both processing methods. In the three-term function, the Gaussian distribution term is related to probability of occurrence of invalid measurements; the power-law term describes the primary behavior of the uncertainty versus the metrics; and a constant expressing the minimum expected uncertainty level for the corresponding methodology, regardless of value of the metrics. The formulas successfully predict the standard and expanded uncertainty coverage close to 68.5% and 95% over the synthetic image sets as well as a 2D stagnation point experiment case using all provide metrics using both SCC and RPC method.

In conclusion, this paper generalizes a framework of models for predicting the expected uncertainty levels for individual velocity measurement in a PIV flow field without the knowledge of local flow condition using only the information contained in the calculated correlation plane. This work continues work establishing the foundations toward the growing understanding of PIV uncertainty estimation.

**Acknowledgments**

The authors wish to acknowledge the support of NSF-IDBR award 1152304 and the NSF/FDA SIR award 1239265. Also PPV would like to thank Barton L Smith, Bernd Wieneke, Andrea Sciacchitano and Doug Neal for the numerous discussions over the past year on various the issues related to the PIV uncertainty quantification.

## 5  References

1. Willert CE, Gharib M. Digital particle image velocimetry. Experiments in Fluids. 1991 1991/01/01;10(4):181-93. English.

2. Adrian RJ. Twenty years of particle image velocimetry. Experiments in Fluids. 2005 2005/08/01;39(2):159-69. English.

3. R M. Mesure de champs de vitesse d' ecoulements fluides par analyse de suites d' images obtenues par diffusion d' un feuillet lumineux: Universite Libre de Bruxelles; 1983.

4. Yao C-S, Adrian RJ. Orthogonal compression and 1-D analysis technique for measurement of 2-D particle displacements in pulsed laser velocimetry. Appl Opt. 1984 1984/06/01;23(11):1687-9.

5. Adrian RJ, Yao C-S. Pulsed laser technique application to liquid and gaseous flows and the scattering power of seed materials. Appl Opt. 1985 1985/01/01;24(1):44-52.

6. Keane RD, Adrian RJ. Optimization of particle image velocimeters. I. Double pulsed systems. Measurement Science and Technology. 1990;1(11):1202.

7. Adrian RJ. Particle-Imaging Techniques for Experimental Fluid Mechanics. Annual Review of Fluid Mechanics. 1991;23(1):261-304.

8. Soloff SM, Adrian RJ, Liu Z-C. Distortion compensation for generalized stereoscopic particle image velocimetry. Measurement Science and Technology. 1997;8(12):1441.

9. Willert C. Stereoscopic digital particle image velocimetry for application in wind tunnel flows. Measurement Science and Technology. 1997;8(12):1465.

10. Scarano F. Iterative image deformation methods in PIV. Measurement Science and Technology. 2002;13(1):R1.

11. Scarano F. Theory of non-isotropic spatial resolution in PIV. Experiments in Fluids. 2003 2003/09/01;35(3):268-77. English.

12. Scarano F. A super-resolution particle image velocimetry interrogation approach by means of velocity second derivatives correlation. Measurement Science and Technology. 2004;15(2):475.

13. Wereley S, Gui L. A correlation-based central difference image correction (CDIC) method and application in a four-roll mill flow PIV measurement. Experiments in Fluids. 2003 2003/01/01;34(1):42-51. English.

14. Wereley ST, Meinhart CD. Second-order accurate particle image velocimetry. Experiments in Fluids. 2001 2001/09/01;31(3):258-68. English.

15. Adrian RJ, Westerweel J. Particle Image Velocimetry: Cambridge University Press; 2010.

16. Timmins B, Wilson B, Smith B, Vlachos P. A method for automatic estimation of instantaneous local uncertainty in particle image velocimetry measurements. Experiments in Fluids. 2012 2012/10/01;53(4):1133-47. English.

17. Sciacchitano A, Wieneke B, Scarano F. PIV uncertainty quantification by image matching. Measurement Science and Technology. 2013;24(4):045302.

18. Keane RD, Adrian RJ. Optimization of particle image velocimeters: II. Multiple pulsed systems. Measurement Science and Technology. 1991;2(10):963.

19. Keane R, Adrian R. Theory of cross-correlation analysis of PIV images. Applied Scientific Research. 1992 1992/07/01;49(3):191-215. English.

20. Hain R, Kähler CJ. Fundamentals of multiframe particle image velocimetry (PIV). Experiments in Fluids. 2007 2007/04/01;42(4):575-87. English.

21. Persoons T, O'Donovan TS. High Dynamic Velocity Range Particle Image Velocimetry Using Multiple Pulse Separation Imaging. Sensors. 2010;11(1):1-18. PubMed PMID: doi:10.3390/s110100001.

22. Charonko JJ, Vlachos PP. Estimation of uncertainty bounds for individual particle image velocimetry measurements from cross-correlation peak ratio. Measurement Science and Technology. 2013;24(6):065301.

23. Adric E, Pavlos PV. Digital particle image velocimetry (DPIV) robust phase correlation. Measurement Science and Technology. 2009;20(5):055401.

24. Eckstein A, Vlachos PP. Assessment of advanced windowing techniques for digital particle image velocimetry (DPIV). Measurement Science and Technology. 2009;20(7):075402.

25. Eckstein A, Charonko J, Vlachos P. Phase correlation processing for DPIV measurements. Experiments in Fluids. 2008 2008/09/01;45(3):485-500. English.

26. Kumar BVKV, Hassebrook L. Performance measures for correlation filters. Appl Opt. 1990 07/10;29(20):2997-3006.

27. Shannon CE. A mathematical theory of communication. SIGMOBILE Mob Comput Commun Rev. 2001;5(1):3-55.

28. Brady MR, Raben SG, Vlachos PP. Methods for Digital Particle Image Sizing (DPIS): Comparisons and improvements. Flow Measurement and Instrumentation. 2009 Sep 09:1-13.

29. Westerweel J, Scarano F. Universal outlier detection for PIV data. Experiments in Fluids. 2005 2005/12/01;39(6):1096-100. English.

30. Stanislas M, Okamoto K, Kähler CJ, Westerweel J. Main results of the Second International PIV Challenge. Experiments in Fluids. 2005 2005/08/01;39(2):170-91. English.

31. Stanislas M, Okamoto K, Kähler CJ, Westerweel J, Scarano F. Main results of the third international PIV Challenge. Experiments in Fluids. 2008 2008/07/01;45(1):27-71. English.